\documentclass[%
 aip,
 amsmath,amssymb,
 reprint,%
]{revtex4-1}

\usepackage{graphicx}
\usepackage{dcolumn}
\usepackage{bm}

\usepackage[utf8]{inputenc}
\usepackage[T1]{fontenc}
\usepackage{bm}
\usepackage{mathptmx}
\usepackage{makecell}  
\usepackage{multirow}
\usepackage{etoolbox}
\usepackage{amsthm,amsmath,amssymb}
\usepackage{mathrsfs}

\makeatletter
\def\@email#1#2{%
 \endgroup
 \patchcmd{\titleblock@produce}
  {\frontmatter@RRAPformat}
  {\frontmatter@RRAPformat{\produce@RRAP{*#1\href{mailto:#2}{#2}}}\frontmatter@RRAPformat}
  {}{}
}%
\makeatother
\begin{document}

\preprint{AIP/123-QED}

\title{Inferring Network Structure with Unobservable Nodes from Time Series Data}
\author{Mengyuan Chen}
\affiliation{School of System Science, Beijing Normal University, No. 19, Xinjiekou Wai Street, Beijing, China, 100875}

\author{Yan Zhang}%
\affiliation{School of System Science, Beijing Normal University, No. 19, Xinjiekou Wai Street, Beijing, China, 100875}

\author{Zhang Zhang}
\affiliation{School of System Science, Beijing Normal University, No. 19, Xinjiekou Wai Street, Beijing, China, 100875}

\author{Lun Du}
\affiliation{Microsoft Research, No. 5 Danling Street, Haidian District, Beijing, China, 10080}

\author{Shuo Wang}
\affiliation{School of System Science, Beijing Normal University, No. 19, Xinjiekou Wai Street, Beijing, China, 100875}

\author{Jiang Zhang}
\affiliation{School of System Science, Beijing Normal University, No. 19, Xinjiekou Wai Street, Beijing, China, 100875}
\email{zhangjiang@bnu.edu.cn}

\date{\today}

\begin{abstract}
Network structures play important roles in social, technological and biological systems. However, the observable nodes and connections in real cases are often incomplete or unavailable due to measurement errors, private protection issues, or other problems. Therefore, inferring the complete network structure is useful for understanding human interactions and complex dynamics. The existing studies have not fully solved the problem of inferring network structure with partial information about connections or nodes. In this paper, we tackle the problem by utilizing time-series data generated by network dynamics. We regard the network inference problem based on dynamical time series data as a problem of minimizing errors for predicting states of observable nodes and proposed a novel data-driven deep learning model called Gumbel-softmax Inference for Network (GIN) to solve the problem under incomplete information. The GIN framework includes three modules: a dynamics learner, a network generator, and an initial state generator to infer the unobservable parts of the network. We implement experiments on artificial and empirical social networks with discrete and continuous dynamics. The experiments show that our method can infer the unknown parts of the structure and the initial states of the observable nodes with up to 90\% accuracy. The accuracy declines linearly with the increase of the fractions of unobservable nodes. Our framework may have wide applications where the network structure is hard to obtain and the time series data is rich. 
\end{abstract}

\maketitle

\begin{quotation}
A complex system is composed of many components that interact with each other. In general, the nodes of a complex network are used to represent the elements in the system, and the edges between nodes are used to represent the interactions, social system, economic system, and biological system are all complex systems. The dependence of nodes in systems is complex. How to mine the network structure of the complex system to help us better understand the behavior of the system has always been a research and challenging problem. This paper aims to study the problem of network structure inference under the absence of system node information -- network completion. The existing research on network completion is to infer the complete network structure in the complex system from the observed information. According to the information obtained, the existing studies can be divided into three categories: one is to use rich node feature information, the other is to use observed structural information, and the third is to make inferences under the scenario in which only temporal sequence information can be observed. Most of the current research assumes that all the node information can be observed, and there is less discussion about the missing information of multiple nodes due to artificial or technical constraints.The contribution of this paper is that it proposes a data-driven, end-to-end method to solve the problem of network completion, and puts forward the application of subgraph matching algorithm to the network completion method, which effectively solves the evaluation problem in network completion.
\end{quotation}

\section{Introduction}

Network structure plays more and more important roles in social, technological, and economic systems
\cite{tran2017community,awate2018geography,chen2013technological,pastore2010heterogeneous}. The connection patterns of a social network determine how fast the opinions or ideas can spread in social media
\cite{i1,i2,centola2010spread}; the structure of a supply chain network between companies influences the safety of the whole market because risk may propagate along with the links
\cite{klibi2012scenario,cimini2015systemic}; the topology of the cooperation network plays a critical role for scientific innovation and individual development for young scientists \cite{newman2001scientific,i3}. Nowadays, many big data analysis, such as recommendation, node importance mining, community clustering, etc., rely on high-quality link data
\cite{ghani2019social}. However, the data of network structure is always incomplete or even unavailable either because measuring binary links is costly or the data of weak ties is unobservable \cite{cimini2015systemic,Kossinet2006effects,anand2018missing}. Therefore, it is urgent to find a way to infer the complete network structure according to non-structural information \cite{squartini2018reconstruction,guimera2009missing}.

Link prediction, as the traditional task in network inference, tries to infer the lost links in network structure according to the linking patterns of existing connections \cite{kunegis2009learning,lu2009similarity}. Although numerous algorithms have been developed to complete the unobservable links of a large network with high accuracy \cite{wang2011human,zhang2018link}, all of these approaches require the complete node information but it is always unavailable in practice \cite{Kossinet2006effects,tran2019deepnc}. Link prediction cannot solve the inference problem under the condition that the network contains unobservable nodes. In real cases, we can either obtain node information of the only partial network or without any information about links \cite{tran2017community,guthke2004dynamic}, as a result, conventional link prediction algorithms cannot work.

Network completion methods have been developed in recent years trying to tackle the problem we discussed above, that is, to infer the missing connections on unobservable nodes according to the linking patterns between observable nodes. The methods can be categorized into the traditional expectation maximum(EM) method \cite{kim2011network,xue2019reconstructing} and graph neural network based methods \cite{wu2019comprehensive,du2018dynamic,du2018galaxy}. As an example of the expectation maximum method \cite{kim2011network}, Kronecker Graph Expectation Maximum (KronEM) algorithm based on Kronecker Graph model \cite{leskovec2010kronecker} can complete the network according to the observable links. Although their algorithm can obtain a relatively high accuracy of recovering missing links, the self-similarity property is required for the underlying network structure as an implicit condition, which is always violated by some networks \cite{leskovec2010kronecker}. On the other hand, with the booming development of deep learning on graphs \cite{wu2019comprehensive,du2018dynamic,du2018galaxy}, researchers applied graph convolution network (GCN) liked models on the network completion problem. Xu et al. proposed a GCN based model, which regards the process of completing a graph as a network growth process, and learns the rules of the growth to complement the full network \cite{xu2019generative}. Tran et al. solved the problem by training a graph generative model to learn the connection patterns among a large set of similar graphs and use these patterns to infer the missing connections \cite{tran2019deepnc}. All of these network completion methods depend on a partially observable network structure because they try to discover the latent patterns of the observable connections and to infer the unknown structures. Nevertheless, in some cases,  the network structures are totally unknown and only some signals of observable nodes can be obtained, such as biological network \cite{geier2007reconstructing} and social network \cite{Kossinet2006effects}. How can we infer the whole network structure without any information on connection patterns?

In fact, time-series data of observable node behaviours can be another important information source \cite{krajec2016highlighting,lin2018variational} which is more or less ignored by previous studies. For example, in an online social network, we can only observe the discrete retweet events between a large set of users, neither their features like sex, education, etc. nor their connection information is unavailable; In a stock market, all the information that we can obtain is the prices of different stocks, the connections between the stocks are unknown. Thus, can we develop a method to infer the network structure according to the time series data representing the observable states of nodes? A large number of methods such as Granger causality \cite{brovelli2004beta,quinn2011estimating}, correlation measurements \cite{stuart2003gene,eguiluz2005scale,barzel2013network}, driving response \cite{timme2007revealing}, compressed sensing \cite{wang2011human,wang2011predicting,wang2011network,shen2014reconstructing}, and graph network \cite{kipf2018neural,zhang2019general} etc., have been proposed for reconstructing network from time series data. However, these methods can only recover the network structure of observed networks and the functional forms of dynamics are always limited by methods. Can we infer all network structure including unobserved part and unobserved node states with partial time series data of observable nodes? Some works try to recover the information of hidden variables by learning the dynamics of a system \cite{ayed2019learning}. However, these works are always based on a grid network and leave the general heterogeneous network structure never discussed. Actually, completing or refining links in a network by node properties and labels is possible as shown in \cite{2019Learning,wang2021graph}, better network can be obtained if we only try to improve the performance of node classification task. Thus, a general framework for reconstructing network topology, completing missing structures, and learning various types of dynamics, from the time series data is possible and necessary. 

In this paper, we develop a universal framework callled Gumbel-softmax Inference
for Network (GIN) to infer the network structure and node information from the time series data with missing nodes. We solve the problem by finding an optimized network structure, a set of appropriate initial states, and an approximator of the network dynamic such that the errors between the observed time series of the observable nodes and the generated time series according to the GIN model is minimized. GIN consists of a network generator, an initial state generator, and a dynamics learner. The network generator is implemented by Gumbel-softmax technique, which can use stochastic gradient descent to differentiably optimize a network. Dynamics learning is realized by a Graph Network (GN) model. This paper is organized as follows: in section \ref{sec:problem} we will formulate the network inference problem in an optimization framework and illustrate the concrete design of each module; the experimental results are discussed in section \ref{sec:results}. We also point out the advantages and weak points of this work which left for future works in section \ref{sec:discussion}.

\section{Problem and Methods}\label{sec:problem}
In this paper, we focus on the inference problem of network structure, initial states, and the network dynamics based on state time series of observable nodes.

\subsection{Problem Definition}
At first, a formal definition is given. Suppose our studied system has an interaction structure described by a binary graph $G = (V, E)$ with an adjacency matrix $A$, where $V = \{v_1, ..., v_N\}$ is the set of nodes, or interchangeably referred to as vertices, and $N$ is the total number of nodes, $E = \{e_{ij}\}$ is the set of edges between the nodes, and $A$ is a binary matrix of which each entry equals $0$ or $1$.

The network dynamic $\mathcal{S}(\psi,A)$ is defined on the graph $G$, where $\psi$ is the dynamical rule which mapping the states of nodes $\bm{x^t}=(x_1^t,x_2^t,\cdot\cdot\cdot,x_N^t)\in \mathcal{R}^{n\times d}$ at time $t$ to the states $\bm{x^{t+1}}$ at time $t+1$, where $\bm{x^{t+1}}=\psi(\bm{x^{t}})$, and $d$ is the dimension of the states. Thus, time series can be generated by the network dynamic $\mathcal{S}$, which are denoted by $\bm{x^{0:T}}=(\bm{x^0},\bm{x^1},\cdot\cdot\cdot,\bm{x^T})$, where $\bm{x^0}$ is the initial state and $T$ is the total time length of the series. 

\begin{figure*}[ht!]
\includegraphics[width=\linewidth]{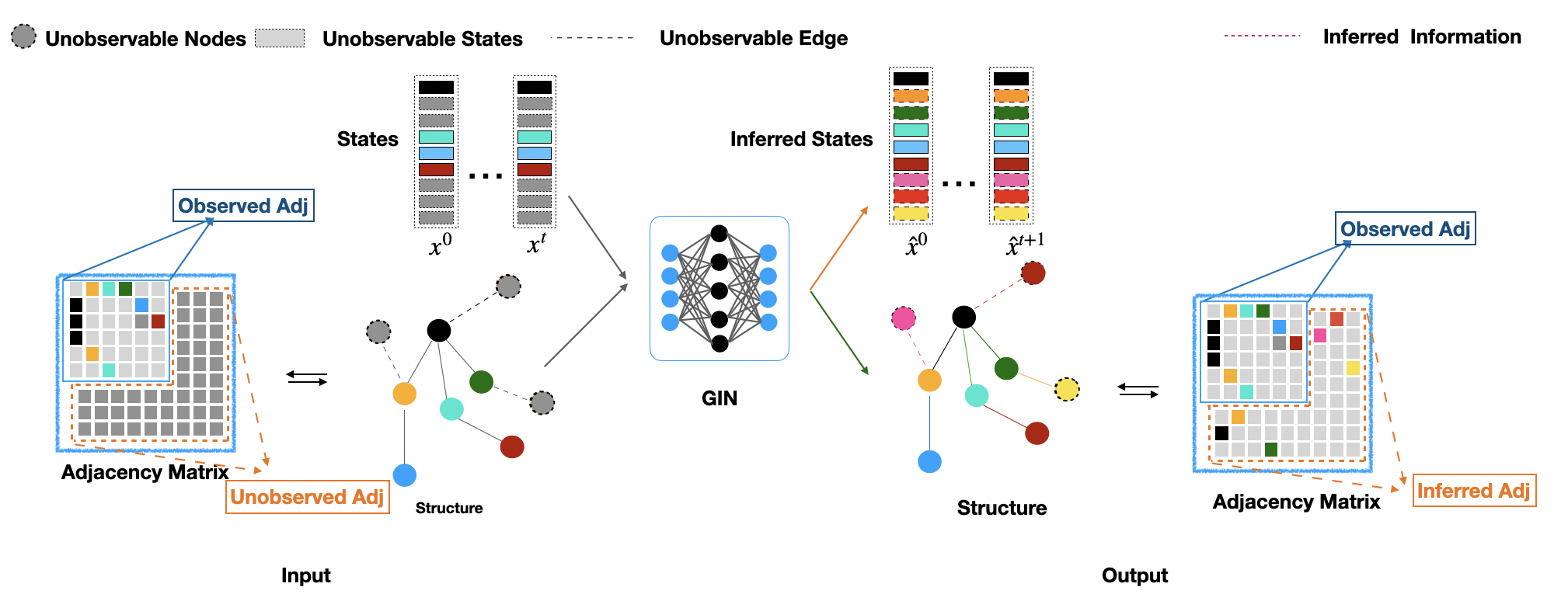}
\caption{The states of some nodes are missing (with dashed box) and only partial network structure can be observed (with bold circles), and the aim is to infer the missing information (the dashed colored circles). In practice, network completion means to infer the “missing” element (with dark grey color) in the adjacency matrix.
}
\label{fig:ncdefinition}
\end{figure*}

However, not all node states can be observed by us. Therefore, the set of nodes $V$ can be divided into two parts: observed nodes $V_o$ and unobserved nodes $V_u$, where $V_o\cup V_u=V$ and $V_0\cap V_u=\emptyset$, and the corresponding state vectors $\bm{x^t}$ can also be decomposed into two parts: $\bm{x^t} = \bm{x}_o^t \bigoplus \bm{x}_u^t$, where $\bigoplus$ is the vector concatenation. Thus, only the partial vector $\bm{x}_o$ can be observed.

Similarly, all of the connections $E$ can also be decomposed into observable connections $E_o$ and unobservable connections $E_u$, and $E_o\cup E_u=E$, and $E_o\cap E_u=\emptyset$. The corresponding adjacency matrix can also be decomposed, $A = A_o\bigoplus A_u$. Where, $A$ is the adjacency matrix of the whole graph $G$, and $A_o$ is the adjacency matrix of the observed connections, and $A_u$ is the one of unobserved connections, $\bigoplus$ is the matrix concatenation after appropriate rearrangement of matrix entries(see the example adjacency matrix in Figure \ref{fig:ncdefinition}, the gray colored entries with the inverted L shape is the unobserved part). 

Note that the two parts of $V$ and the two parts of $E$ do not necessarily have a corresponding relationship. Figure \ref{fig:ncdefinition} shows a general case with the overlap between observable nodes and unobservable connections.

Then, our task is to infer all the unobserved information including node dynamical rules $\psi$, unobserved connections $A_u$, and unobserved node states $\bm{x}_u^t$ according to all the known information including time series of observed nodes $\bm{x}_o^{0:T}$ and observed connections $A_o$.

\subsection{Optimization Problem Formulation}
The network inference problem can be formulated as an optimization problem that finds a set of optimal parameters $\alpha,\beta,\gamma$, to minimize the error value between the state estimation value and the ground-truth, which is the objective function Equation \ref{loss}.
\begin{equation}
 \min\limits_{\alpha,\beta,\gamma} L(\alpha,\beta,\gamma) = \sum_{t=1}^{T}D\left(\bm{x}_o^t, \hat{\bm{x}}_o^t(\alpha,\beta,\gamma)\right) + \lambda ||\hat{A}(\beta)||  
 \label{loss}
\end{equation}

such that:

\begin{equation}
    \bm{\hat{x}_o^t} \bigoplus \bm{\hat{x}_u^t}= \hat{\psi_{\alpha}}\left(\bm{x_o^{t-1}} \bigoplus \bm{\hat{x}_u^{t-1}},\hat{A}(\beta)\right), \forall{t}>1,
\end{equation}

\begin{equation}
    \bm{\hat{x}_u^0}=\rho(\gamma)
    \label{initial states}
\end{equation}

Here, $D(x,y)$ is a measure of the closeness between the state $\bm{x_o^t}$ and $\bm{\hat{x}_o^t(\alpha,\beta,\gamma)}$, and it can be a cross-entropy measure when the states are binary or Mean Absolute Error(MAE) when the states are real numbers. $\hat{\psi_{\alpha}}(\cdot)$ is a dynamical rule approximator to estimate $\psi$ parameterized by $\alpha$. Notice that, to estimate a better state at time step $t$, we use the information from real data of the observable nodes $\bm{x_o^{t-1}}$ at step $t-1$. In this way, we can iteratively apply $\hat{\psi_{\alpha}}$ to the estimated state in the previous time to obtain an estimated evolutionary trajectory $(\bm{\hat{x}^1},\bm{\hat{x}^2},\cdot\cdot\cdot,\bm{\hat{x}^T})$ starting from the state $\bm{\hat{x}^0}=\bm{x_o^0}\bigoplus\bm{\hat{x}_u^0}$. And this trajectory must be similar to the real trajectory $(\bm{x^1},\bm{x^2},\cdot\cdot\cdot,\bm{x^T})$. Further, $\hat{A}(\beta)$ is the estimate of the adjacency matrix $A$ with the parameter $\beta$. And $\hat{\bm{x}}^0=\rho(\gamma)\in \mathcal{R}^{M\times d}$ is an estimate of the initial states of unknown nodes which are parameterized by $\gamma$. The second term in Equation \ref{loss} is the structural loss which can compel the generated adjacency matrix to be sparse, and $\lambda>0$ is the parameter to balance the relative importance between the structural loss and the prediction error. 

We have converted the network inference problem into an optimization problem, but this is a very general framework in which concrete implementation should be given.

\subsection{Network Inference Framework}

\subsubsection{Implementation Gumbel-softmax Inference for Network}
To implement the framework mentioned in the previous paragraph, we propose a concrete implementation called Gumbel-softmax Inference for Network (GIN). GIN is composed of a network generator based on gumbel softmax technique and a dynamics learner based on graph network technique \cite{zhang2019general,zhang2021automated}.

\begin{figure*}[ht!]
\centering
\includegraphics[scale=0.35]{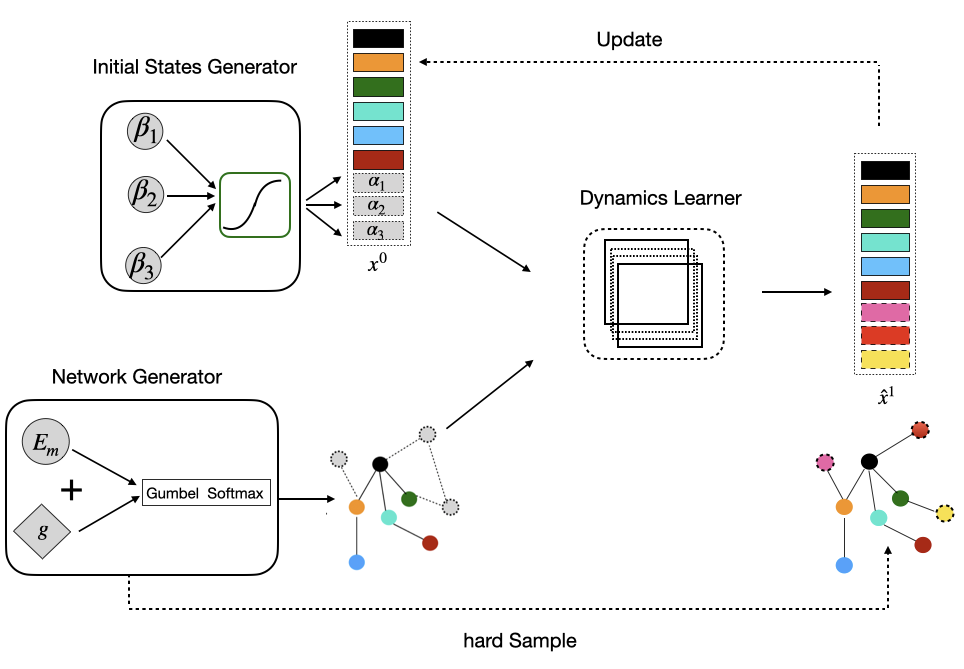}
\caption{The modules of the GIN model. At first, the network structure and the initial state can be generated by the network generator and initial state generator modules, respectively. After that, they are input into the dynamics learner to output the predicted value of the node states at the next time step.}
\label{fig:ncframework}
\end{figure*}

The Framework is shown in Figure \ref{fig:ncframework}. The inputs of our model are the states \bm{$x_o^t$} of observable nodes and the observable adjacency matrix $A_o$. Correspondingly, the outputs are the complete network structure and the future states of all nodes.

At first, the candidate network is generated by a series of gumbel softmax sampling processes parameterized by a matrix $\beta_{N\times N}$, that is,

\begin{equation}
A_{ij} = \frac{\exp((\log(\beta_{ij})+\xi_{ij})/\tau)}{\exp(\log(\beta_{ij})+\xi_{ij})/\tau))+\exp(\log(\beta_{ij})+\xi_{ij}')/\tau))},
\label{eq:gumbel}
\end{equation}
where $\beta_{ij}$ is the probability of the connection between node $i$ and node $j$, and $\xi_{ij}$ is the i.i.d random variable of the standard Gumbel distribution, and $\tau$ is the temperature parameter. When $\tau$ goes to zero, $A_{ij}$ will converge to $0$ or $1$. Equation \ref{eq:gumbel} simulates the sampling process of generating $A_{ij}$ with the probability $\xi_{ij}$, however, it is differentiable such that it can be adjusted by the gradient descent method.

The second module of GIN is the initial state generator. Because the iteration of the dynamics leaner $\psi_{\alpha}$ requires the initial state of all nodes, however, the states of unobserved nodes are missing. We generate the initial states of these unobserved nodes with the initial state generator $\rho(\gamma)$ parameterized by $\gamma$. Here, $\rho$ can be simply an identity function (this is equivalent to sample initial state by $\gamma$ directly) or a function to specify the value boundary of the initial states (for example, a sigmoid function can limit the initial states to the interval $(0,1)$). 

\begin{figure*}
\centering
\includegraphics[scale=0.45]{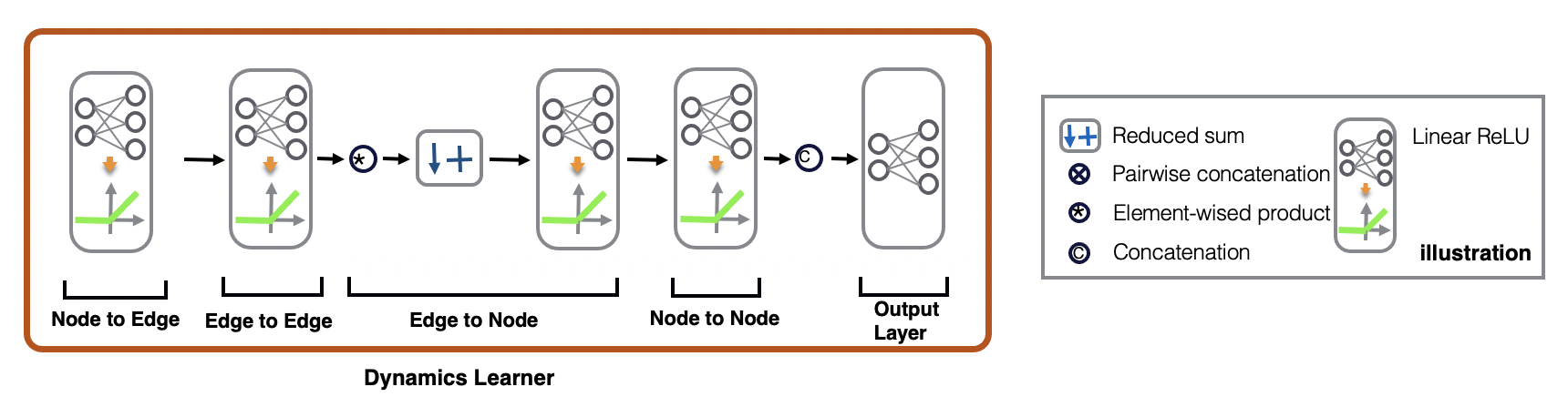}
\caption{ The dynamics learner consists of four parts: (1) Node to Edge: aggregating the original information of nodes to form representations of edges; (2) Edge to Edge: update the edge representations; (3) Edge to Node: aggregate all information on neighbouring edges of each node to form a new feature vector of the current node; (4) node to node: update the node representations; (5) Output: finally, concatenate the node representations and the input state vectors of node i to feed into a feedforward network, and output the prediction of the next state all nodes.
}
\label{fig:dynamics_learner}
\end{figure*}

Third, when the candidate network and initial state are generated, they will be fed into the dynamics leaner module. We assume that the dynamics $\psi$ is node symmetric, therefore, we can use a graph network $\hat{\psi}_{\alpha}$ parameterized by $\alpha$ to implement the dynamics learner as shown in Figure \ref{fig:dynamics_learner}.

We train and update the parameters of the three modules in each epoch. After the predicted states of the unobservable nodes are obtained, the loss function can be calculated by comparing the predicted states and the real ones. And we implement the back-propagation algorithm to obtain the gradient values and update the parameters of the three modules simultaneously. We layout the pseudo-codes of GIN in Algorithm 1 to show the details.

\begin{table}[ht!]
\begin{tabular}{p{0.5\textwidth}}
 \hline
\textbf{Algorithm 1} : GIN algorithm  \\ [0.5ex] 
 \hline
 1 \textbf{Input}: the observed adjacency matrix $A_o$ if have;\\ \hspace*{1.2cm}the time series of all or partial nodes $\bm{x_o}^{0:T}$; \\
 \hspace*{1.2cm}the number of observed nodes $N_o$ if have. \\
 \hspace*{1.2cm}the number of unobserved nodes $N_u$ if have. \\
 2 \textbf{Output}: the predict adjacency matrix $\hat{A}$;\\ 
 \hspace*{1.4cm}the initial states of the unobserved nodes and the predict\\ \hspace*{1.4cm}states of all nodes
 $\bm{\hat{x}} = \{{\bm{\hat{x}_u^0},\bm{\hat{x}_u^{1:T+1}},\bm{\hat{x}_o^{1:T+1}}}\}$.  \\
   \# Initialization \\
 4 Initialize Dynamics Learner parameters $\alpha$ \\
 5 Initialize Initial States Generator parameters $\gamma$\\ 
 6 Initialize Network Generator parameters $\beta(N_u)$ \\
  \# Training \\
 7 \textbf{for} each epoch \textbf{do} \\
 8 \hspace*{0.4cm} Get initial states of unobserved nodes: $\bm{\hat{x}_u^0} = \rho(\gamma)$\\
 9 \hspace*{0.5cm} Sample unobserved adjacency matrix:\\ \hspace*{0.8cm}  $\hat{A}_u=$Network Generator($\beta)$\\
 10 \hspace*{0.3cm} $\hat{A} \leftarrow  (A_o\bigoplus \hat{A}_u)$ \\
 11 \hspace*{0.3cm} \textbf{for} t=0,$\cdots$,$T$ do \\
 12 \hspace*{0.9cm} Concatenate nodal states: $\bm{\hat{x}^t} \leftarrow (\bm{x_o^t} \bigoplus \bm{\hat{x}_u^t)}$ \\
 13 \hspace*{0.9cm} \textbf{for} i=1,$\cdots$,$N_o$ do \\
 14 \hspace*{1.4cm} 
 $\bm{\hat{x}_o^t[i]} \bigoplus \bm{\hat{x}_u^t[i]} \leftarrow$ Dynamics\hspace*{0.1cm} Learner\hspace*{0.1cm}$(\hat{A}[i],\bm{\hat{x}^{0}},\alpha)$ \\
 15 \hspace*{1.4cm} loss $\leftarrow$ Compute\hspace*{0.1cm} Loss\hspace*{0.1cm}$(\{\bm{x_o^t[i]}\},\{\bm{\hat{x}_o^t[i]}\})$ \\
 16  \hspace*{1.4cm} update $\gamma$, $\beta$,\ $\alpha$\ with the gradient of loss \\
 17  \hspace*{0.9cm} \textbf{end}\\
 18  \hspace*{0.3cm} \textbf{end} \\
 19  \textbf{end} \\
 
\hline
\end{tabular}
\label{table:algorithm 1}
\end{table}

\subsubsection{Graph Matching Problem in the Evaluation Process}
After training the GIN framework, we need to evaluate the effect of the network inference. However, it is difficult to evaluate a network completion algorithm in real life because the missing connections are unknown. Our strategy is to find a real network as the ground truth, cut off some of the nodes and edges for testing, and then compare the inferred part of the algorithm with the real network.

However, a new problem, graph matching between the inferred sub-graph and the real one arises during this comparison, because the unobservable nodes between the inference and the ground truth should be aligned before evaluating.

To solve the problem, we can search for all node alignment and find the best one. Here, the best alignment means that each node pair in the alignment has the most similar neighbor relationship with each known node. Thus, the graph matching problem can be formulated as another optimization problem:

\begin{equation}
    \min\limits_{p\in\mathscr{P}({N_u})}{\parallel A - (I_{N_o} \bigoplus P)\hat{A}(I_{N_o} \bigoplus P)^T\parallel}_F^2
    \label{eq:graph_matching}
\end{equation}

Where, $A$ and $\hat{A}$ represent the adjacency matrices of the ground truth and the inference, respectively. $P$ is a permutation matrix(node alignment) with size $n\times n$, and $\mathscr{P}({N_u})$ is the set of all possible permutation matrices with size $N_u$. $I_{N_o}$ is an identity matrix with size $N_o$. The symbol $\bigoplus$ represents the concatenation of matrices. Therefore, $A(I_{N_o}\bigoplus P)^T$ means the rearrangement of the rows and columns of the matrix $A$ with the observable part unchanged.  Thus, formula \ref{eq:graph_matching} means to find an optimized permutation $P$ of the unobservable nodes such that the adjacency matrices between the inference and the ground truth can be as similar as possible.

However, there are $N_u!$ possible permutations such that finding an optimized permutation by brute force searching is impossible. Therefore, we use Seed Graph Matching (SGM) algorithm to solve this NP-hard problem \cite{2019Seeded}.
 
At first, the objective function in Equation \ref{eq:graph_matching} can be expanded as:

\begin{equation}
\begin{aligned}
\label{eq:matching_factor}
&{\parallel A - (I_{N_o} \bigoplus P)\hat{A}(I_{N_o} \bigoplus P^T)\parallel}_F^2=\\
&{\parallel A \parallel}_F^2 + {\parallel \hat{A}\parallel}_F^2 - 2\cdot  {\rm Tr} \left(A^T(I_{N_o} \bigoplus P)\hat{A}(I_{N_o} \bigoplus P^T)\right)
\end{aligned}
\end{equation}

where $\parallel$. $\parallel_F$ is the Frobenius norm on matrices. Then, the minimization problem can be further simplified to the maximization problem:
\begin{equation}
\max \limits_{p\in \mathcal{D}({N_u})}
 J(P)=Tr\left(A^T(I_{N_o} \bigoplus P)\hat{A}(I_{N_o} \bigoplus P^T)\right).
\end{equation}
Nevertheless, this optimization problem is also hard to solve because $P$ is a permutation matrix with binary entries. We then relax the problem by allowing the matrix $P$ to be a doubly stochastic matrix such that the value range of each entry can be extended to the interval $[0,1]$ as suggested by the SGM algorithm. After that, the conjugated optimization method can be used to optimize $J(P)$. The details can be referred to \cite{2019Seeded}. As reported by \cite{2019Seeded}, when the similarity of the two graphs is more than 90\% and the number of matched nodes is more than 15, the matching accuracy of this algorithm can be more than 90\%. In our experiment, the parameters are the same as those in \cite{2019Seeded}.

\section{Experimental Results}\label{sec:results}
Our framework and algorithms are universal because they can be applied to any network structure and any type of time series data such as continuous or binary states. 
\subsection{Data Set}
We test our method on both synthetic and empirical social networks.
\subsubsection{Synthetic Network}
We generate synthetic networks by well-known network models such as: ER(Erdos-Renyi network ) \cite{bollobas2001random}, WS(Small world network) \cite{watts1998collective} and BA(Scale-free network) \cite{albert2002statistical}. Next, we summarized the parameter settings for generating synthetic networks.

\begin{itemize}
\item[-]\textbf{ER}. In the ER model, nodes are linked to each other with the probability $p=0.04, 0.013, 0.01$ for the networks with the number of nodes as 100, 300 and 1000, respectively. 

\item[-]\textbf{WS}. The WS model can be used to generate locally clustered networks. In our setting, we at first connect 4 closest neighbors on a ring, and we rewire a link randomly with 0.3 probability.
\item[-]\textbf{BA}. The BA model can simulate the scale-free property of real networks. The preferential attachment rule is used to grow a BA network. At the beginning, there are $m_0=20$ nodes have been existed as the seeds. Then, a new node is added which will connect $k=2$ existing nodes with preferential attachment rule.
\end{itemize}

\subsubsection{Empirical Social Networks}
We also select 6 real social networks with different structures as the representatives of the empirical networks. Except for Dorm, the connections of these networks are undirected, which means that the transmission of information between nodes is mutual rather than one-way. 


The basic structural parameters of the three social networks are shown in Table \ref{network}

\begin{table}
\caption{\label{network}Network parameter of social network}
\begin{ruledtabular}
\begin{tabular}{cccccc}
   Network & $N$\footnote{number of nodes} & $E$\footnote{number of edges} &$\langle k\rangle$ \footnote{average degree}& $\langle C\rangle$\footnote{average clustering coefficient}& $\langle r\rangle$\footnote{degree-degree correlation coefficient}\\
  \hline
    {Karate} &34 &78 & 4 & 0.5706 & -0.4756 \\
  {Dolphins} &62 &159 & 5 & 0.2589 & -0.0435 \\
  {Email-partial} &143 &623 & 8 & 0.4339 & -0.01953 \\
  {Dorm} &217 &2672 & 24 & 0.399 & 0.1195 \\
  {Email} &1133 &10902 & 4 & 0.220  &0.0782\\
  {Blog} &1224 &19025 & 15 & 0.210& -0.2200 \\
\end{tabular}
\end{ruledtabular}
\end{table}

\subsection{Time Series Data of Network Dynamics}
A large number of time series data is required to implement our approach, however, it is hard to obtain from the real scenario because the problems of privacy and measurement are concerned. Therefore, we use synthetic time series data generated from the network dynamics instead of real data. Two different types of time series data (binary and real-valued) are tested because both types can be processed by our approach.

The first type of time series is binary which simulates the process of opinion spreading on a social network. We use the well known Voter model to simulate the opinion dynamic on network. The Voter model is introduced by Richard A. Holley and Thomas M. Liggett in 1975 \cite{holley1975ergodic,wang2012dynamics}. Suppose there are $N$ interacting agents connected to form a network. Initially, each agent has a distinct ``opinion'' represented by $x_i^t=\{0,1\}$. At each time $t$, any agent $i$ will have a chance to change his ''opinion'', and the probability to adopt the opinion is determined by the relative fraction of the same opinion in all of $i$'s neighbors.

The second type of dynamics on the social network has real-valued state. This models the cases that the psychological states of different people (e.g., the expectation price of a stock or a commodity) can influence each other via social connections.

We choose the chaotic network dynamic Coupled Mapping Network (CMN) as our candidate to generate the time series. A Coupled map network (CMN) model is a network dynamic with discrete time and continuous state which is proposed by Kaneko in 1992
\cite{kaneko1992overview}. Each element on a network
consists of a logistic map coupled to their neighbors, this can be written as 
\begin{equation}
    x^{t+1}_i = (1-\epsilon)f(x^{t}_i)+\frac{\epsilon}{|N_i|}\sum_{j\in {N_i}}f(x^{t}_j)
\end{equation}
where $x^{t}_i\in [0,1]$ is the state of node $i$ at time $t$, $N_i$ represents node $i$'s neighbors, $\epsilon\in (0,+\infty)$ is the coupling constant which can tune the system behavior, and the local map $f(x)$ is the logistic map:
\begin{equation}
    f(x) = x(1-x).
\end{equation}
In our experiments, we set $\epsilon=3.5$
\subsection{Data Preparation}
We have evolved discrete Voter dynamics on the synthetic network and real social networks, and continuous CMN dynamics on synthetic networks. In order to generate time series data for each node, we first generate $s$ initial states for each node. In each initial state, we evolve $T$ time steps forward through the dynamic function. In the process of model training, $t$ time-step state information is used. The number of training data we derived from the dynamics of CMN is $S= s \times T /t$ , the number of data evolving from the Voter dynamics is $S =(T-t+1) \times s$. In all experiments, we set $t = 2$. The $T$ values of CMN and Voter dynamics are 100 and 51, respectively. On networks of different sizes and tasks, we generate different size data sets by adjusting $s$.

On network completion tasks, including part of the network structure is known and no network structure, we set $s$ to 50 in a synthetic network of 100 nodes, and generated 2.5k data sets with a time step of 2, and in 300 nodes set $s$ to 400 on the synthetic network and generate 20k data sets based on CMN dynamics. For Voter dynamics, we generated 5k and 15k volume data sets with steps of 100 and 300 on the synthesized network of 100 nodes and 300 nodes, respectively.  In particular, we used 15K of data on a 100-node ER network. On the karate, Dolphin, and Email networks, $s$ is 20, 200, and 300,  and 1k, 10k, and 15k data set are generated respectively. The division ratio of the training set, test set and validation set is 5:1:1.

For the network reconstruction task, on the continuous data set generated on the CMN dynamics, we generate 12k, 10k and 60K data on 10,100 and 1000 nodes of the WS network respectively. The ratio of the training set, test set and verification set is 10:1:1. We set the value at 10 and 100 respectively. 12k, 10k, and 60k data are generated on the WS network of 1000 nodes. The division ratio of training set, test set and verification set is 10:1:1. For Voter dynamics, the division ratio of the data set is 5:1:1. The $s$ values of the WS network with 10, 100, and 1000 nodes are 400, 200, and 2000, and the data volume is 20k, 10k, and 100k. On the real network Dorm, Blog $s$ is 1200, 3200, data volume is 60k, 160k respectively.

In appendix, we summarize the amount of experimental data. For each experiment, we repeat three times and then calculate the average value to report.




\subsection{Experimental Setup}
To implement the experiments, several representative tasks are set. The parameters in different situations are also given in this subsection.
\subsubsection{Tasks}
Our framework GIN can be applied to any case with or without observable nodes and partial networks. Without lose of generality, three specific tasks are designed as follows:
\begin{itemize}
\item[-] \textbf{Network Completion with Partial Structural Information}: In this task, we randomly select a fraction of nodes as unobservable nodes, and we remove the corresponding time series and the corresponding entries of the adjacency matrix (all connections related to the unobservable nodes). And the incomplete data is fed on the framework to ask GIN to infer the unknown adjacency matrix and the unobservable initial node states.
\item[-] \textbf{Network Completion without Structural Information}: In this task, we need to randomly select some nodes as unobservable, and remove the corresponding time series to feed into GIN. Nevertheless, different from the previous task, the partial network is never known for the framework.
\item[-] \textbf{Network Reconstruction}: In this task, we need to reconstruct all links from the observed time series, and all nodes are observable.
\end{itemize}
\subsubsection{Parameter Settings}
To achieve good results in the tasks mentioned, we need to set up the parameters in GIN.

In general, we set parameters in all experiments as follows:

(1) In the dynamics learner module, we use a 4-layered MLP as a node shared function as shown in Figure \ref{fig:dynamics_learner}. The activation function of each layer is ReLU. The model parameter $\alpha$ is randomly initialized. We use 32 hidden units in each layer in most cases. However, it is 64 on the Voter model.

(2) In the initial state learning module of the network completion task, discrete and continuous datasets have different initial state generation methods. The nodes' states are usually coded by one-hot vectors on the data set generated by Voter. For the data sets of CMN, We use sigmoid as the function $\rho$ to map the parameters $\gamma$ into the interval $[0,1]$

(3) In the network generator module, because the adjacency matrix constructed is symmetric, we just randomly generate $N(N-1)/2$ elements of the upper triangular part of the adjacency matrix via the gumbel-softmax sampling process parameterized by $\beta$ which are sampled by a normal distribution $N(0,0.1)$, and we add up the transposed triangular matrix to obtain a complete symmetric adjacency matrix. For example, in Karate Network, we generate only 33 * 17 parameters, which is the number of triangular elements on the adjacency matrix except for diagonals.

The above three modules are optimized by using Adam optimizer, the learning rates of (1) to (3) module are set to $0.004$, $0.1$, and $0.001$, respectively. The learning rate of the state leaner is higher than that of the other two modules. This is because states have too many parameters, and it is easy to fall into local optimums.  For each epoch, we randomly select 1,024 samples to train and can have better results after training about 500 epochs.

The structural loss parameter $\lambda$ is set to be zero in all tasks except the network reconstruction task which is set to be 0.0001.
\subsubsection{Performance on Network Completion with Partial Structural Information}
First, we test the performance of GIN on the network completion task with partial structural information. According to our investigation so far, we do not find a method that can be applied directly to solve the network completion task based on time series data, thus, we do not compare it to other models. We carry out the experiment of the GIN model on network completion on 100-scale and 300-scale networks, where the percents of missing nodes are 10.

\begin{table*}
\caption{\label{tab:completion}Network Completion Performance on different networks and dynamics}
\begin{ruledtabular}
\begin{tabular}{cccccccc}
Dynamics                    & $N-N_u$     & Network & Unobs-AUC & Unobs-ACC(net) & Unobs-TPR & Unobs-FPR & \makecell{Obs-ACC/MSE} \\
\hline
\multirow{9}{*}{Binary}     & \multirow{3}{*}{100-10} & ER      & 0.8355$\pm0.103$&0.8933$\pm0.0152$&0.6241&0.1002&81.01\%      \\

 &                         & WS    & 0.9218$\pm0.005$&0.8333$\pm0.006$&0.8869&0.1691&82.07\% \\
 &                         & BA            & 0.9233$\pm0.006$&0.8133$\pm0.023$&0.8993&0.1844&82.19\%  \\ \cline{2-8}
 & \multirow{3}{*}{300-20} & ER            & 0.9550$\pm0.009$&0.9600$\pm0.000$&0.6385&0.0387&77.29\%
 \\
 &                         & WS            & 0.9583$\pm0.005$&0.9400$\pm0.006$&0.7672&0.0539&79.33\%  \\
 &                         & BA            & 0.9387$\pm0.004$&0.9567$\pm0.006$&0.6054&0.0397&79.31\% \\ \cline{2-8}
 & 34-3                    & Karate        & 0.8241$\pm0.05$&0.6800$\pm0.144$&0.7847&0.4042&87.93\%  \\
 & 62-6                    & Dolphins    & 0.8742$\pm0.011$&0.7900$\pm0.010$&0.8366&0.2157&83.50\%  \\
 & 143-14                  & Email-partial & 0.7915$\pm0.010$&0.8550$\pm0.007$&0.4964&0.1238&76.52\%  \\
\hline
\multirow{6}{*}{Continuous} & \multirow{3}{*}{100-10} & ER      & 0.9839$\pm0.008$     & 0.9500$\pm0.008$          & 0.8723      & 0.0441      & 8.03E-06    \\
 &                         & WS            & 0.9463$\pm0.032$ & 0.9033$\pm0.052$ & 0.8509 & 0.1185 & 9.68E-06   \\
 &                         & BA            & 0.9282$\pm0.003$ & 0.9333$\pm0.0205$ & 0.8282 & 0.0580 & 7.99E-06 
 \\\cline{2-8}
 & \multirow{3}{*}{300-20} & ER            & 0.9902$\pm0.008$ & 0.9533$\pm0.023$ & 0.9579 & 0.0464 & 1.42E-06 
  \\
 &                         & WS            & 0.9463$\pm0.082$ & 0.9033$\pm0.137$ & 0.8509 & 0.1185 & 9.68E-06 
 \\
 &                         & BA            & 0.9286$\pm0.017$ & 0.9467$\pm0.032$ & 0.8259 & 0.0487 & 4.13E-06 
 \\ 
\end{tabular}
\end{ruledtabular}
\end{table*}

In table \ref{tab:completion}, the first numbers in the N-$N_u$ column represent the total network size, and the second numbers are the numbers of nodes being removed. For example, 100-10 indicates that we remove 10 nodes from a whole network with 100 nodes. The indicator of unobs-AUC refers to the Area Under the ROC Curve of the prediction for the unobservable part of the network. Also, we display some details about the results, such as the unobs-ACC(net) which is the proportion of elements that correctly estimated adjacency matrix $\hat{A}$, and the values of the True Positive Rate (TPR), False Positive Rate (FPR) of the unobservable part. The obs-ACC/MSE (states) is the accuracy of the predicted future states of observed nodes for the binary dynamic (Voter), and it is the mean square error for the continuous dynamic (CMN).

GIN can perform well on both tasks of network inference and observed node states prediction. The accuracy of the state of observable recovery is over 77\% for the binary data set. And the relative error rate is close to 0 in the continuous data sets. Besides, our model maintains a stable accuracy on different network structures. Table \ref{tab:completion}  demonstrates that the accuracy of the binary data set is lower than that of the CMN data sets especially on the FPR results and also has a larger deviation of the dynamic state prediction value.  It can be explained by the fact that the information propagation process of the Voter dynamic is stochastic, however, the CMN dynamic is deterministic.

\begin{figure*}
    \centering
    \includegraphics[width=\linewidth]{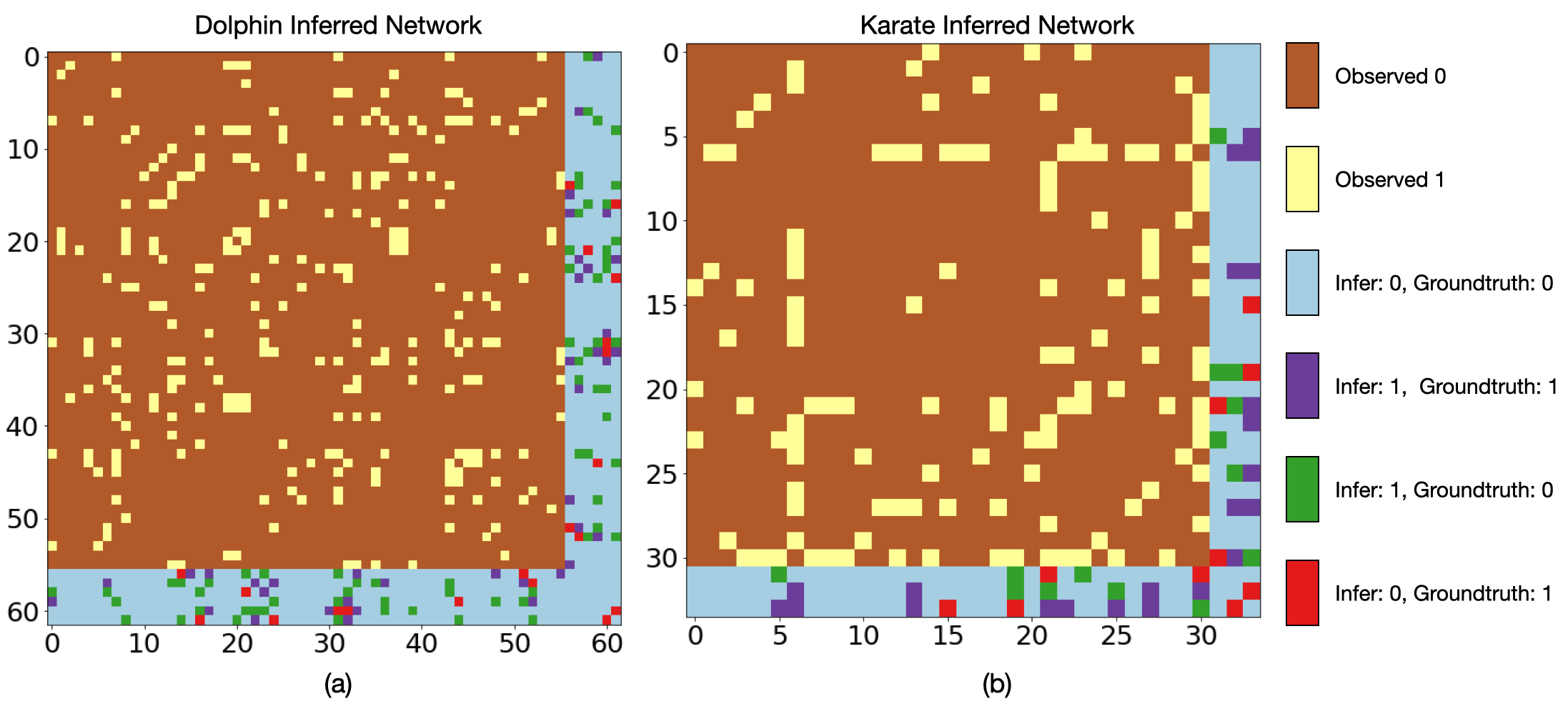}
    \caption{Contrast matrices of the adjacency matrices between the inference and the ground truth for Dolphin network (a), and the Karate network (b). The invert L-shaped part in the figure has four colors, among which blue is the True Positive element, purple is the True Negative element, green is False True element, and red is the False Negative element.}
    \label{fig:inferadjmat}
\end{figure*}

We show a heat map Figure \ref{fig:inferadjmat} comparing the inferred adjacency matrix graph with the real adjacency matrix to describe the effect of our completion more clearly. We set a reasonable threshold to turn the inferred probability adjacency matrix into an adjacency matrix that is either zero or one, then stack it on top of ground truth and the following heat map is plotted finally.  Both the brown and yellow boxes in the figure represent the observed local network structure and the remaining invert "L" shape is the unobservable part in which the green and red squares represent elements that are incorrectly inferred. On the Dolphin network, red and green elements account for about 20\%, and on the Karate network, the proportion of red and green elements is about 24\%. Notice that, all correctly labeled matrix units are concentrated in the same row or column, which shows that our GIN algorithm can infer that some observed nodes have some connections with some unobservable nodes. However, it is hard to know which unobservable nodes are.

We also do experiments on three empirical social networks, where the percent of missing nodes is set to be 10\%. Table \ref{tab:completion} includes the metrics to evaluate the structural accuracy and the state recovery accuracy. GIN achieves over 80\% accuracy, which is less than in the synthetic network. One of the possible reasons is that the real social network is denser than the synthetic network, which increases the uncertainty of state transition compared to a sparse network. Our model obtains about 80\% accuracy in extrapolating the state values, which means that our model learned the dynamics of the nodes from discrete dynamics. Both in the inference of missing structure and known nodes' states, our model has achieved high inference accuracy on the empirical social network. 

\begin{figure}[ht!]
\centering
\includegraphics[scale=0.5]{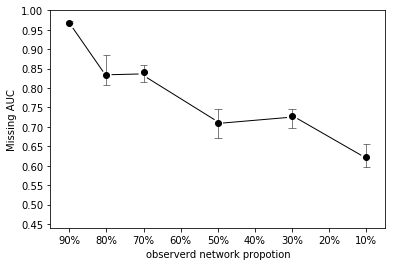}
\caption{AUC of unobservable network decreases with the proportion of observed nodes.}
\label{network completion part}
\end{figure}

Generally speaking, inference accuracy should decrease with the increase of the proportion of unobservable nodes in the network completion problem. Thus, we investigate how fast the accuracy decrease with the missing proportion increase from 10\% to 90\%.  
Figure \ref{network completion part} shows the effect of the accuracy of network inference on the proportion of observed nodes. It can be seen that the AUC value decreases in a linearly way approximately. That means, each time the percentage of observed nodes increases by 10\%, the AUC decreases by 0.05. 

Through the above experiments, we found that GIN can infer network structure information with a higher accuracy rate. In addition, the effect of the inference is affected by the observed information. When the ratio of the observed information is less than 50\%, it is difficult to infer the complete dynamic system accurately.

\subsubsection{Network Completion without structural information}
\begin{figure}[ht!]
\centering
\includegraphics[scale=0.35]{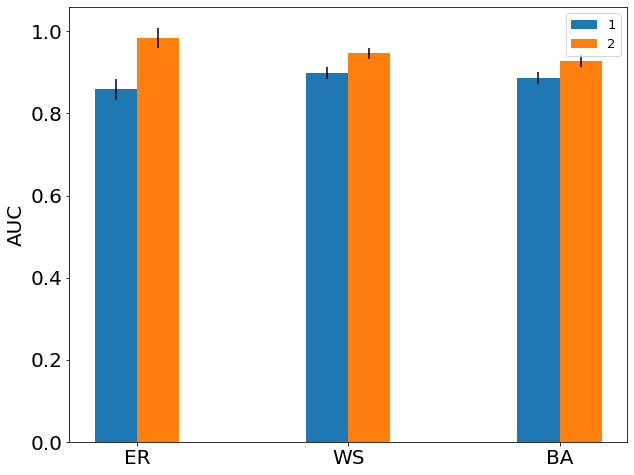}
\label{network completion framework}
\caption{Comparison of the performance of GIN under the network completion task with (1) or without (2) network structure information.}
\label{vs completion}
\end{figure}

In the previous task, partial network structures can be known. However, in some real cases, we can only know the time series of the observed nodes. Therefore, we test the task of network completion without structural information in this experiment. To finish this task, we first use GIN to reconstruct the observable nodes and then make the completion of the whole network according to the reconstructed network. We outline our results by comparing the network completion task with or without network structure between observed nodes in Figure \ref{vs completion}. It can be seen that the performance of task 2 is significantly better than that of task 1. However, the performance of task 1 is also very good, and the AUC is above 80\%.

Further, we show the AUCs of network inference for different parts as shown in Table \ref{without structural}. It can be seen that on different social networks, the AUC of the whole network has reached above 0.9. While, the AUCs of the unobservable part are 0.6, 0.6, and 0.8 on Karate, Email, and Dolphin networks, respectively. Reconstruction AUC represents the accuracy of the network reconstruction task for the links between observable nodes. We also compare the inferred network and the ground truth on a set of representative statistical indicators, and the results are similar as shown in the table \ref{tab: structural properties}.

In summary, our framework can also perform certain completion work when structural information is missing.

\begin{table*}
\caption{\label{without structural}Network completion without structural information}
\begin{ruledtabular}
\begin{tabular}{lccccc}

Dynamics    & Network       & Nodes-Missing nodes & Missing AUC & Reconstruction AUC & All  AUC \\ \hline
\multirow{6}{*}{Discrete} & Karate        & 34-3                  & 0.7602      & 0.9930          & 0.9524   \\
                            & Dolphins      & 62-6                  & 0.8237      & 0.9989          & 0.9766   \\
                            & Email-partial & 143-14                & 0.5899      & 0.9819          & 0.9231   \\
                            & ER            & 100-10                & 0.8923      & 0.9908          & 0.9850   \\
                            & WS            & 100-10                & 0.8622      & 0.9883          & 0.9957   \\
                            & BA            & 100-10                & 0.9189      & 0.9956          & 0.9875   \\ \hline
\multirow{3}{*}{Binary}     & ER            & 100-10                & 0.8585      & 0.9931          & 0.9717   \\
                            & WS            & 100-10                & 0.8979      & 0.9943          & 0.9808   \\
                            & BA            & 100-10                & 0.8863      & 0.9795          & 0.9776   \\ 
\end{tabular}
\end{ruledtabular}
\end{table*}

\begin{table}
\caption{\label{tab: structural properties}Comparison of the statistical properties for Dolphin network}
\begin{ruledtabular}
\begin{tabular}{ccc}
Statistical characteristics & GIN & Real \\ 
\hline
Average Degree & 5.067 & 5.129 \\ 
Graph Distance & 3.26 & 3.357  \\ 
Graph Density & 0.086 & 0.084  \\ 
Clustering Coefficient & 0.294 & 0.303 \\

\end{tabular}
\end{ruledtabular}
\end{table}

\subsubsection{Network Reconstruction}
On the task of network reconstruction, we compare with the state-of-the-art methods such as neural relational inference model (NRI) \cite{kipf2018neural}. 
and the algorithm for revealing network interactions (ARNI) \cite{casadiego2017model}. In addition, we also compared with two traditional methods, namely mutual information method \cite{donges2009backbone} and partial correlation  method \cite{mccabe2020netrd}.

\begin{itemize}
\item[-] \textbf{NRI}(Neural Relational Inference Model) applies a variational auto-encoder method to learn the underlying interaction graph and the complex system dynamics from the observational dynamical data. We ran the NRI Model by using the settings which are consistent with the original paper in \cite{kipf2018neural}.
\item[-] \textbf{ARNI}(Algorithm for Revealing Network Interactions) is a state-of-the-art method of network reconstruction task by regressing the gradient information of node state with the state in the previous time step \cite{casadiego2017model}.
\item[-] \textbf{Pcorr \& MI} (Partial Correlation \& Mutual Information) are all measures of correlation between node states. Partial correlation refers to the process of removing the influence of the third variable when two variables are simultaneously related to the third variable and only analyzing the degree of correlation between the other two variables. Mutual information (MI) is an information theoretic measure of the correlation of two variables. On network reconstruction, the two methods can be used to measure the similarity between the time series of two nodes. The less the similarity, the greater the probability that the two nodes connect.
\end{itemize}
\begin{table*}
\caption{\label{network reconstruction}Network reconstruction performance on different networks and dynamics}
\begin{ruledtabular}
\begin{tabular}{cccccccccc}
\multirow{2}{*}{Dynamics} & \multirow{2}{*}{Nodes} & \multicolumn{2}{c}{GIN} & MI & PCorr & \multicolumn{2}{c}{ARNI} & \multicolumn{2}{c}{NRI} \\ \cline{3-10} 
                        &           & AUC    & ACC/MSE & AUC    & AUC   & AUC & ACC/MSE & AUC    & ACC/MSE \\ \hline
\multirow{5}{*}{Binary} & WS-10     & 1      & 0.9463     & 0.525  & -     & -   & -          & 0.5037 & 0.9062     \\
                        & WS-100    & 0.9961 & 0.7914     & 0.508  & -     & -   & -          & -      & -          \\
                        & WS-1000   & 0.9996 & 0.6623     & 0.547  & -     & -   & -          & -      & -          \\
                        & Dorm-217  & 0.6918 & 0.9999     & 0.5219 & -     & -   & -          & -      & -          \\
                        & Blog-1224 & 0.6366 & 0.9715     & 0.4995 & -     & -   & -          & -      & -          \\ \hline
Discrete               & WS-10     & 1      & 3.31E-04   & 0.6875 & 0.785 & 1   & 2.35E-09   & 0.9997 & 8.40E-08   \\
                        & WS-100    & 0.9987 & 1.48E-06   & 0.571  & 0.613 & -   & -          & -      & -          \\
                        & WS-1000   & 0.9995 & 2.92E-06   & 0.567  & -     & -   & -          & -      & -          \\ 
\end{tabular}
\end{ruledtabular}
\end{table*}

In Table \ref{network reconstruction}, we show the performances of our model on the network reconstruction task. AUCs of GIN can reach above 99\% on WS small world networks with different sizes. However, ARNI and NRI models can work on small networks with continuous dynamics. Our Framework can also handle large networks with more than a thousand of nodes while maintaining performance.


\subsubsection{Utility analysis of seed graph matching algorithm}

In order to analyze the effectiveness of our proposed evaluation algorithm, we conducted ablation experiments on the existing model by deleting the seed graph matching algorithm module, and the inferred adjacency matrix is directly compared with the real adjacency matrix. In addition, we also show the accuracy of matching by SGM between a randomly generated adjacency matrix and the real adjacency matrix.

We conducted experiments on a synthetic network with 100 nodes. 
\begin{table*}
\caption{\label{table:sgmabla}Comparison table of the effect of SGM algorithm module in GIN model}
\begin{ruledtabular}
\begin{tabular}{cccc}
   Network & AUC GIN with SGM  & AUC GIN without SGM & Random AUC with SGM\\
  \hline
 ER(100-10) &{0.9839}{$\pm0.008$} &  0.7662{$\pm0.062$} & 0.6323{$\pm0.004$}\\
 BA (100-10) &{0.9463}{$\pm0.032$} &0.8540{$\pm0.019$} & 0.6885{$\pm0.008$}  \\
 WS(100-10) &{0.9282}{$\pm0.025$} &0.6268{$\pm0.024$} &0.6519{$\pm0.011$} \\
\end{tabular}
\end{ruledtabular}
\end{table*}
The specific results are shown in table\ref{table:sgmabla}. The column "AUC (GIN without SGM)" shows the AUC value inferred from the adjacency matrix of the missing network structure without the Seeded Graph Matching module (Seeded Graph Matching, referred to as SGM), and uses SGM to match the randomly generated adjacency matrix. It is clear that GIN with SGM can get highest value of AUC on network completion. The result reflects the Network Completion -The optimal effect can be achieved only when GGN is coupled with the SGM evaluation algorithm, which shows the effectiveness of the SGM algorithm.
\subsubsection{Computational complexity analysis}
Training the the neural network requires time, and the time complexity increases with network size. To test how the time complexity depends on network size, we conduct experiments on a GPU of Tesla V100(16G) on WS small-world networks with 20,50,100, 200,300 nodes. And on the networks, 10\% of nodes are unknown.

\begin{figure}
    \centering
    \includegraphics[width=0.8\linewidth]{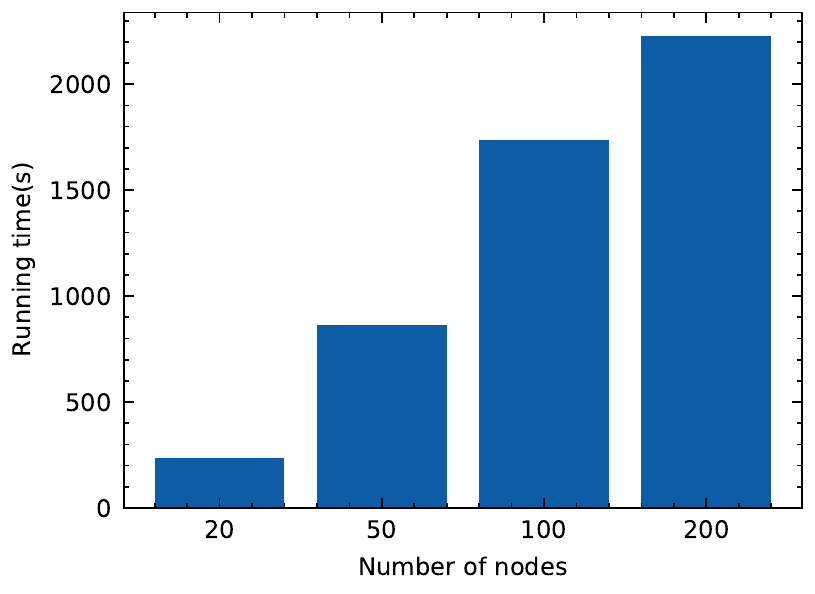}
    \caption{The running time of GIN increases with the network size and the number of missing nodes.}
    \label{fig:time}
\end{figure}

Fig \ref {fig:time} shows the curve of time complexity. Training the model requires an hour when the network size is not exceed 200. However, the time complexity increases dramatically on a network with 300 nodes due to the increase of the requirement on time series data,  and it takes almost five hours.

\section{Discussion}\label{sec:discussion}
In the paper, we discuss the problem of network inference with few unobservable nodes, and we propose a universal framework to solve the problem. First, we formulate the network inference problem based on time series data under an optimization framework. Second, we design GIN model by integrating three modules: a Gumbel-softmax based network generator, a graph network based dynamics leaner, and an initial state generator. Third, we apply GIN on two vastly different types of time series data. We then reported the performances of GIN framework on three different network inference tasks, and GIN can work while on both network inference and initial state inference.

The benefits of our framework include its lightweight design, high accuracy, and universality to different network structures and dynamics. By using the Gumbel-softmax based network generator and the initial state generator, we can simply set the unknown elements to be learnable parameters, the design can be generalized in more cases. We test our framework on three different kinds of network inference tasks, and it can achieve relatively good results on all the tasks. And the results are robust and universal for different network structures and dynamics.

However, there are still many aspects that can be improved in our current work. For example, our model have a high accuracy in networks where the missing percent is less than 10\%. With the missing percent increase, the number of unobserved nodes that have more than one degree of separation from an observed node will increases significantly which results in a major loss in the performance of the GIN model. The performance of the initial state generator can be improved on a larger space of node state. Besides, the amount of data needed for network inference is relatively large, the reduction for data requirement should be solved in the future. Furthermore, we hope that the accuracy of network completion can be further improved by increasing the performance of the initial state generator. 

In future works, we hope to combine the information of node states information and network structural information to infer unknown network structures. The dynamics learner also can be generalized to non-Markovian dynamical processes.

\begin{acknowledgments}
The research is supported by the National Natural Science Foundation of China(NSFC) under the grant numbers 61673070.
\end{acknowledgments}

\section*{Data Availability Statement}

The data that support the findings of this study are available from the corresponding author upon reasonable request.

\section*{Appendices}
\subsection{Data number on different task}
As shown in table \ref{Data number on different task}, we summarize the  details about experimental data.

\begin{table*}
\caption{Data on different task}
\begin{ruledtabular}
\begin{tabular}{ccccc}
Task &	dynamics & Network & Number of initial states s	& Number of datasets S
\\ \hline
\multirow{8}{*}{Network Completion Task}     & \multirow{2}{*}{CMN} & 100-node synthetic &	50 &	2.5k \\
& & 300-node synthetic &	400 & 20k \\ \cline{2-5}
& \multirow{6}{*}{Voter} & 100-node WS/BA	& 100 &	5k \\
& & 100-node ER	& 300 &	15k \\
& & 300-node synthetic	& 300 &	15k  \\
& & Karate &	20 &	1k \\
& & Dolphin & 200 & 10k \\
& & Email-partial &	300 & 15k \\
\cline {2-5}
\multirow{8}{*}{Network Reconstruction Task } & \multirow{3}{*}{CMN} & 10-node &	240 &	12k \\
& & 100-node &	200	& 10k \\
& & 1000-node &	1200	& 60k \\ \cline{2-5}
 & \multirow{5}{*}{Voter} & 10-node &	400	& 20k \\
& & 100-node	& 200	& 10k \\
& & 1000-node & 2000	& 100k \\
& & Dorm & 1200 & 60k  \\
& & Blog & 3200 & 160k 
\label{Data number on different task}
\end{tabular}
\end{ruledtabular}
\end{table*}

\subsection{Convergence of the algorithm}
\begin{figure}
    \centering
    \includegraphics[width=0.8\linewidth]{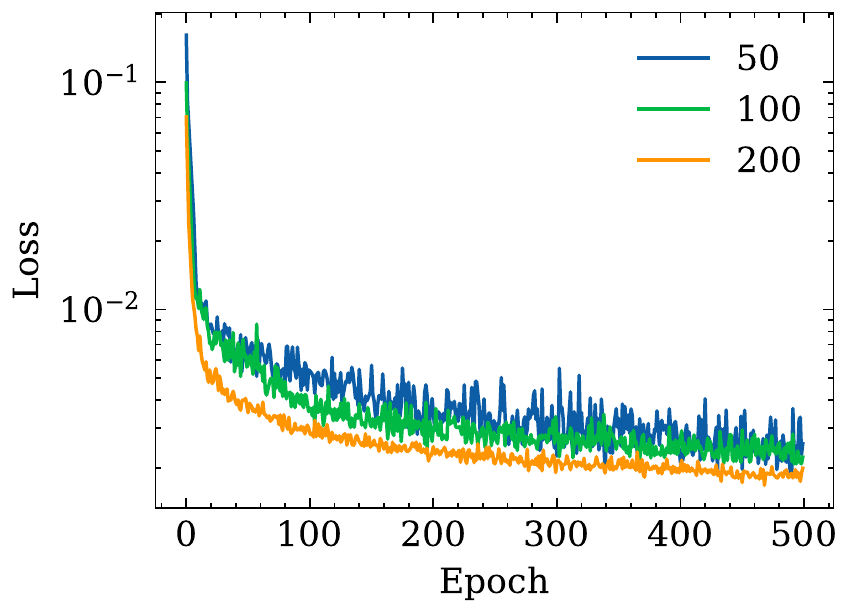}
    \caption{The convergence scale with increasing network size. As the training epoch increased, loss(MAE) with logarithm  shows a clear downward trend. At different node scales, the models converge as the training epoch increased.}
    \label{fig:loss}
\end{figure}

Figure \ref{fig:loss} shows the convergence scale with increasing network sizes. As the size of nodes increases, loss (MAE) is always maintained at a relatively stable level, which proves that the GIN algorithm has good convergence.

\section*{REFERENCES}
\nocite{*}
\bibliography{aipsamp}

\end{document}